\documentclass[useAMS,usenatbib]{mn2e}

\usepackage{graphicx}
\usepackage{color}

\newcommand{\be}{\begin{equation}}
\newcommand{\ee}{\end{equation}}
\newcommand{\bea}{\begin{eqnarray}}
\newcommand{\nn}{\nonumber}
\newcommand{\eea}{\end{eqnarray}}

\title[QPOs as probes of compact objects structure]
{What can quasi-periodic oscillations tell us about the structure
of the corresponding compact objects?}
\author[George Pappas]{George Pappas$^{1}$\thanks{E-mail:
gpappas@phys.uoa.gr} \\
$^{1}$Section of Astrophysics, Astronomy, and
Mechanics, Department of Physics, University of Athens,
\\Panepistimiopolis Zografos GR15783, Athens, Greece}

\begin{document}

\date{Accepted 2012 February 23. Received 2012 February 22; in original form 2011 December 21}

\pagerange{\pageref{firstpage}--\pageref{lastpage}} \pubyear{2012}

\maketitle

\label{firstpage}

\begin{abstract}

We show how one can estimate the multipole moments of the
space-time, assuming that the quasi-periodic modulations of the
X-ray flux (quasi-periodic oscillations), observed from accreting
neutron stars or black holes, are due to orbital and precession
frequencies (relativistic precession model). The precession
frequencies $\Omega_{\rho}$ and $\Omega_z$ can be expressed as
expansions on the orbital frequency $\Omega$, in which the moments
enter the coefficients in a prescribed form. Thus, observations
can be fitted to these expressions in order to evaluate the
moments. If the compact object is a neutron star, constrains can
be imposed on the equation of state. The same analysis can be used
for black holes as a test for the validity of the no-hair theorem.
Alternatively, instead of fitting for the moments, observations
can be matched to frequencies calculated from analytic models that
are produced so as to correspond to realistic neutron stars
described by various equations of state. Observations can thus be
used to constrain the equation of state and possibly other
physical parameters (mass, rotation, quadrupole, etc.) Some
distinctive features of the frequencies, which become evident by
using the analytic models, are discussed.
\end{abstract}


\section{Introduction}

Low mass X-ray binaries (LMXBs) open a window to some extreme
physics. On the one hand the accretion in these systems takes
place in the region around a compact object (such as a black hole
or a neutron star) where there are strong gravity effects. On the
other hand, when the accretion takes place around a neutron star,
the properties of the orbital motion depend on the structure of
the neutron star itself. Thus studying these systems could provide
insight on the properties of gravity in regions where it is yet
untested and at the same time help us explore the properties of
matter in such extreme conditions as the ones in the interior of
neutron stars.

One of the properties that LMXBs often display, are quasi-periodic
oscillations (QPOs) in their X-ray flux. This variability appears
in a variety of behaviors and is usually organized in various
groupings, such as the high-frequency phenomena, the hectohertz
QPOs, or the low-frequency complex. The aforementioned
categorization is based not only on the frequency range but also
on the characteristics that the observed frequencies display which
depend on the type and the state of the source. For example one
can observe, (i) in the low-frequency complex low-frequency QPOs
with $\nu$ values in the range of 0.01 up to 100 Hz or (ii) in the
case of high-frequency phenomena, neutron star kHz QPOs (often
observed in pairs), called twin kHz QPOs, which have been observed
up to $\nu=1258\, \mathrm{Hz}$ (\cite{jonker}). On the other hand
there are high-frequency QPOs which are categorized as black hole
high-frequency QPOs and appear to be different than their neutron
star counterparts. Finally, the QPOs, apart from the previously
mentioned differentiation exhibit also correlations between the
various frequency components, both in neutron stars as well as in
black holes (for a review see \cite{lamb};\cite{derKlis}).

To explain the QPOs, various types of mechanisms have been
proposed. These are: (i) the Beat-frequency models, where one
assumes that there is some ``beating" of an orbital frequency by
the spin frequency of the central object, (ii) the relativistic
precession models, where the QPOs are associated with the orbital
motion and the periastron or nodal precession of a particular
orbit, (iii) the relativistic resonance models, where a type of
resonance between the orbital and the epicyclic frequencies is
assumed wherever they have simple integer ratios, and finally (iv)
the preferred radii models, where some mechanism chooses a
particular radius. These models generally assume the geodesic or
almost geodesic orbits of the fluid elements in the accretion disc
to be the source of the observed frequencies (see e.g.
\cite{derKlis}), while there are also models in which the
frequencies are produced from oscillatory modes of the entire disc
(see e.g. \cite{Rezzolla}). In one way or another all of these
models use the properties of the orbits around the compact object
onto which the accretion takes place. In our discussion we will
refer to the models that assume that the QPOs are caused by the
frequencies associated with the orbital motion of the material in
the accretion disc such as the relativistic precession models (see
\cite{stella}), but our analysis has relevance to the other models
as well.

In the relativistic precession models it is assumed that there is
some structure or inhomogeneity in the accretion disc (i.e. a hot
blob) that follows a keplerian or geodesic orbit which is almost
circular and almost at the equatorial plane. Such an orbit has an
orbital frequency $\Omega$ and precession frequencies,
$\Omega_{\rho}$ of radial perturbations and $\Omega_z$ of
azimuthal (vertical) perturbations. We note here that in the
following, while we will assume that such a structure is
incorporated in an accretion disc, the orbit of the structure is
such that it can be considered to be geodesic, i.e. all forces
other than gravity can be ignored. This can be considered to be a
good approximation in the case of radiatively efficient thin
discs.

The most commonly used assumption for describing the
aforementioned frequencies is that the background space-time is
either Schwarzschild or Kerr or that it is well approximated by
one or the other. A more general assumption for the background
space-time would be to assume a stationary and axisymmetric
space-time as the one given by the \cite{papapetrou} line element,

\be \label{pap} ds^2=-f\left(dt-\omega d\phi\right)^2+
      f^{-1}\left[ e^{2\gamma} \left( d\rho^2+dz^2 \right)+
      \rho^2 d\phi^2 \right].
\ee

\noindent Such a space-time can be described with the help of the
Ernst potential (\cite{ernst1,ernst2}) and the metric functions
can be associated to the multipole moments defined by
\cite{geroch,hansen} and \cite{fodor}. Thus one can express the
rotational and epicyclic frequencies as functions of the multipole
moments of the space-time. In particular, it was demonstrated by
\cite{ryan} that the quantities $\Omega_{\rho}/\Omega$ and
$\Omega_z/\Omega$ can be expressed as series expansions on
$\Omega$ where the coefficients depend on the multipole moments of
the underlying space-time. Although Ryan's work was intended to be
applied on the analysis of gravitational waves from the 
adiabatic inspiral of a compact object onto a super-massive
central object (extreme mass ratio inspirals, EMRIs), the same
principle can be applied in the case of X-ray observations from
accretion discs (in LMXBs), assuming that the orbits in the disc
are almost circular and almost equatorial.

The expressions given by Ryan are asymptotic expansions that do
not give a very good description for the regions close to the
innermost stable circular orbit (ISCO). In that case, one could
make use of the several proposed analytic space-times that give a
faithful description of the geometry around neutron stars (see
\cite{Stuart,BertiSterg,Pachon,Pappas2,Teich}) and use them to
work in the strong gravity region. These analytic solutions are
constructed by using the first few multipole moments of the
space-time which are associated to the properties of the compact
object. Thus the assumption of such solutions could relate the
frequencies of the QPOs, to the moments of the compact object and
give a more accurate description for the QPOs than the typically
used geometries of Schwarzschild and Kerr. In our analysis, we
will use the two-soliton analytic solution (see
\cite{twosoliton,Pappas2}) which is constructed from the first
four multipole moments of the neutron star, i.e., the mass ($M$),
the angular momentum ($J$), the mass quadrupole ($M_2$) and the
spin octupole ($S_3$).

The benefit of this approach is that instead of relying on
gravitational waves (that are at the moment eluding us) to get
information for the central compact object, one can rely on
electromagnetic observations which are currently abundant. Thus,
an algorithm that would produce estimates for the multipole
moments of the space-time in low mass X-ray binaries could be used
to probe either the equation of state for neutron stars or the
no-hair theorem for black holes. Following this point of view,
there is currently a programme put forward by \cite{psaltis} on
testing gravity with observations in the electromagnetic spectrum.

Our aim in this work is to demonstrate how one could estimate the
first few space-time moments by using the series expansions of the
frequencies (presented in section 2) and fitting them to QPOs (in
section 4). We also present templates for the relation between the
various frequencies (in section 5), which are constructed with the
help of an analytic solution of the Einstein field equations which
has been chosen appropriately so as to describe faithfully the
geometry around neutron stars built from realistic EOSs (briefly
presented in section 3). Using these templates we show how one
could constrain the EOS for matter of higher than nuclear
densities that exists inside neutron stars (section 6).

\section{The frequencies as functions of\\* the multipole moments}

In this section we will briefly present the expressions that
describe the orbital motion of a test particle in an axisymmetric
space-time near the equatorial plane and outline the algorithm
that associates the precession frequencies with the orbital
frequency and the multipole moments as it was developed by
\cite{ryan}.

A stationary and axisymmetric space-time can generally be written
in the form of the Papapetrou line element given in Eq.
(\ref{pap}), where $f,\;\omega$ and $\gamma$ are functions of the
Weyl-Papapetrou coordinates ($\rho,\,z$). In that case, the metric
components are:

\[ g_{tt}=-f,\quad g_{t\phi}=f\omega,\quad
g_{\phi\phi}=f^{-1}\rho^2-f\omega^2,\nn \]
\[g_{\rho\rho}=g_{zz}=
f^{-1} e^{2\gamma}, \nn \]

\noindent where the radial coordinate is defined as $\rho^2=
g_{t\phi}^2-g_{tt}g_{\phi\phi}$. The orbit of a particle is
usually calculated from the conservation of energy, the
conservation of the z component of the angular momentum (where z
is the direction of the symmetry axis) and the normalization of
the four-velocity. Thus, the two conservation equations are:

\be
\label{E}\frac{E}{\mu}=-g_{tt}\left(\frac{dt}{d\tau}\right)-g_{t\phi}\left(\frac{d\phi}{d\tau}\right),\ee
\be\label{Lz}\frac{L_z}{\mu}=g_{t\phi}\left(\frac{dt}{d\tau}\right)+g_{\phi\phi}\left(\frac{d\phi}{d\tau}\right),
\ee

\noindent while the normalization of the four-velocity gives:

\bea
-1&=&g_{tt}\left(\frac{dt}{d\tau}\right)^2+2g_{t\phi}\left(\frac{dt}{d\tau}\right)\left(\frac{d\phi}{d\tau}\right)\nn\\
\label{velocity}&&+
g_{\phi\phi}\left(\frac{d\phi}{d\tau}\right)^2+g_{\rho\rho}\left(\frac{d\rho}{d\tau}\right)^2+g_{zz}\left(\frac{dz}{d\tau}\right)^2.
\eea

If the orbit of the particle is on the equatorial plane, then one
could define an effective potential from the equation

\be \label{Veff}
-g_{\rho\rho}\left(\frac{d\rho}{d\tau}\right)^2=1-
\frac{\tilde{E}^2g_{\phi\phi}+2\tilde{E}\tilde{L}_zg_{t\phi}+\tilde{L}_z^2g_{tt}}{\rho^2}\equiv
V(\rho), \ee

\noindent where $\tilde{E}$ and $\tilde{L}_z$ are the conserved
energy and angular momentum per unit rest mass, respectively, and
$V(\rho)$ is the effective potential. From this equation one sees
that the conditions for circular orbits are, $d\rho/d\tau=0$ and
$d^2\rho/d\tau^2=0$, which are equivalent to the conditions for a
local minimum of the effective potential: $V(\rho)=0,\quad
dV(\rho)/d\rho=0$. For the circular motion, one can also derive
the orbital frequency,

 \be
\Omega\equiv\frac{d\phi}{dt}=\frac{-g_{t\phi,\rho}+\sqrt{(g_{t\phi,\rho})^2-g_{tt,\rho}g_{\phi\phi,\rho}}}{g_{\phi\phi,\rho}}.
\ee

In case the orbit is not exactly circular or not exactly on the
equatorial plane, one can use expressions (\ref{E}-\ref{velocity})
to derive the equation for the perturbed circular motion and from
these the epicyclic or precession frequencies $\Omega_{\rho}$ and
$\Omega_z$. The expression for the frequencies is

 \bea
\Omega_{\alpha}&=&\Omega- \left[-\frac{g^{\alpha\alpha}}{2} \left((g_{tt}+g_{t\phi}\Omega)^2 \left(\frac{g_{\phi\phi}}{\rho^2}\right)_{,\alpha\alpha}\right.\right.\nn\\
&&-2(g_{tt}+g_{t\phi}\Omega)(g_{t\phi}+g_{\phi\phi}\Omega)\left(\frac{g_{t\phi}}{\rho^2}\right)_{,\alpha\alpha}\\
&&\left.\left.+
(g_{t\phi}+g_{\phi\phi}\Omega)^2\left(\frac{g_{tt}}{\rho^2}\right)_{,\alpha\alpha}\right)\right]^{1/2},\nn
\eea

\noindent where if we set $\alpha\rightarrow\rho$, we obtain the
periastron precession and if we set $\alpha\rightarrow z$, we get
the nodal precession. Thus, we have the expressions for the
orbital frequency and the precession frequencies of an almost
circular and almost equatorial orbit.

These expressions depend on the metric functions and their first
and second derivatives with respect to the coordinates $\rho$ and
$z$, evaluated on the equatorial plane. As described in
\cite{ryan}, these functions can be expressed as power series in
$1/\rho$. So, by replacing these expressions in the definition of
the orbital frequency we obtain

\be \Omega=(M/\rho^3)^{1/2}(1+\textnormal{series in}~
\rho^{-1/2}). \ee

\noindent This series can be inverted to take the form

 \be 1/\rho =
(\Omega^2/M)^{1/3}(1+\textnormal{series in}~ \Omega^{1/3}).\ee

If we use the above expression to eliminate the $1/\rho$
dependence of the precession frequencies, we end up with a series
with respect to $\upsilon=(M\Omega)^{1/3}$ (where, in the
Newtonian limit, $\upsilon$ is the particle's orbital velocity),

\bea
\label{ryan1} \frac{\Omega_{\rho}}{\Omega}=\sum_{n=2}^{\infty}R_n\upsilon^n \\
\frac{\Omega_{z}}{\Omega}=\sum_{n=3}^{\infty}Z_n\upsilon^n. \eea
The coefficients of the series, $R_n$ and $Z_n$, depend on the
Geroch-Hansen-Fodor multipole moments of the space-time. For
example, the first few non-vanishing coefficients are

\[R_2=3,\quad\!\! R_3=-4\frac{S_1}{M^2},\quad\!\!
R_4=\left(\frac{9}{2}-\frac{3M_2}{2M^3}\right), \ldots\nn\]
\[Z_3=2\frac{S_1}{M^2},\quad\!\! Z_4=\frac{3M_2}{2M^3},\quad\!\!
Z_6=\left(7\frac{S_1^2}{M^4}+3\frac{M_2}{M^3}\right), \ldots\]

\noindent where $M$ and $M_2$ are the mass and the quadrupole
moment, respectively, and $S_1$ is the angular momentum. These
expressions extended up to terms of $O(\upsilon^{10})$ can be
found in \cite{ryan}.

We should point out that in our analysis we are using the term
``frequency" for the $\Omega$ values which are actually angular
velocities while in the literature the observations of QPOs are
given as ``cycles/time" ($\nu$ values). The two are related as
$\Omega=2\pi\nu$. Whenever we refer to the $\nu$ frequencies it
will be clearly stated so. All quantities are calculated in
geometric units (km) except for the frequencies which are usually
given in kHz. The conversion from $\textnormal{km}^{-1}$ to kHz is
performed by multiplying by $299.79$, i.e.
$1\,\textnormal{km}^{-1}=299.79\,\textnormal{kHz}$. In these units
1 solar mass ($M_{\odot}$) is 1.477 km.

\section{The space-time around neutron stars}

In this section we will present the method that we have used to
model the space-time around neutron stars. First we briefly
discuss the procedure used for generating analytic space-times and
then we talk about the choice of specific neutron star models.

\subsection{The analytic space-time}

As we have said previously, a stationary and axially symmetric
space-time can be described by the line element (\ref{pap}). In
that case the Einstein field equations in vacuum reduce to the
Ernst equation (see \cite{ernst1}),

\be \label{ErnstE}
Re(\mathcal{E})\nabla^2\mathcal{E}=\nabla\mathcal{E}\cdot\nabla\mathcal{E},
\ee

\noindent where the Ernst potential $\mathcal{E},$ is a complex
function of the metric functions.

A general procedure for generating solutions of the Ernst equation
has been developed by \cite{sib1,SibManko,manko2} and
\cite{twosoliton}. The solution of the Ernst equation is produced
from a choice of the Ernst potential along the axis of symmetry of
the form

\be \mathcal{E}(\rho=0,z)=e(z)=\frac{P(z)}{R(z)}, \ee

\noindent where the functions $P(z), R(z)$ are polynomials of
order $n$ with complex coefficients in general. In Ruiz et al.
(1995), one can find a detailed description of the procedure for
generating the solutions by using the parameters of the
polynomials in the Ernst potential. It is shown in Ruiz et al.
(1995) that the metric functions are finally expressed in terms of
some determinants.

The vacuum two-soliton solution (constructed by \cite{twosoliton})
is a special case of the previous general axisymmetric solution
that is obtained from the ansatz (see also \cite{Sotiriou})

\be \label{2soliton}
e(z)=\frac{(z-M-ia)(z+ib)-k}{(z+M-ia)(z+ib)-k},\ee

\noindent where all the parameters are real, while the parameters
$M$ and $a$ are the mass and the reduced angular momentum $J/M$
respectively. For this analytic solution, the first five mass and
mass-current moments of the corresponding space-time are

\[  M_0=M,\quad M_1=0,\quad M_2=-(a^2-k)M,\quad
M_3=0,\] \[ M_4=\left( a^4 - (3a^2 -2ab + b^2)k + k^2
+\frac{1}{7}(2a^2-k)M^2\right) M\] \[ S_0=0,\quad J=aM,\quad
S_2=0,\] \be\label{moments} S_3=- (a^3 -(2a - b)k)M  ,\quad S_4=0.
\ee

\noindent As it can be seen, the first four moments are linearly
related to the parameters of the Ernst potential and thus if they
are specified, one can generate an analytic space-time with
exactly these first four moments. Specifically, the parameter $k$
is uniquely fixed by the quadrupole $M_2$ and the parameter $b$ is
uniquely fixed by the spin octupole $S_3$, while $M$ and $a$ are
the mass and the angular momentum per mass, respectively.

The two-soliton solution has been thoroughly studied and has been
shown to be a faithful description of the exterior of neutron
stars and captures all the characteristics of the corresponding
numerical space-times (see \cite{Pappas2}, and for more details
\cite{Pappas3}).

\begin{table}
 \centering
  \caption{Frequencies calculated for the models constructed with the L EOS. The mass
  and the radii (circumferential radii at the equator and at ISCO) are given in km.
  The frequencies are given in kHz. The frequencies
  are computed at the ISCO except for the cases that the ISCO falls bellow the surface of the
  star; then the corresponding frequencies are computed on the surface and are indicated
  by an asterisk. We note that the frequencies shown are angular frequencies, i.e. $\Omega=2\pi\nu$.}
\begin{tabular}{|c|c|c|c|c|c|c|}\hline
$M$ & $j$ & $R_s$  & $R_{isco}$  & $\Omega$  & $\Omega_{\rho}$  & $\Omega_z$  \\
(km) &  & (km) & (km) & (kHz) & (kHz)& (kHz) \\ \hline
 2.080 & 0.    & 14.96 & 12.48 & 7.47* & 4.42* & 0* \\
 2.071 & 0.194 & 15.18 & 11.76 & 7.26* & 3.78* & 0.062* \\
 2.075 & 0.324 & 15.61 & 11.98 & 7.00* & 3.47* & 0.026* \\
 2.080 & 0.417 & 16.06 & 12.32 & 6.74* & 3.25* & -0.019* \\
 2.083 & 0.483 & 16.49 & 12.63 & 6.51* & 3.06* & -0.054* \\
 2.087 & 0.543 & 16.97 & 12.94 & 6.26* & 2.87* & -0.085* \\
 2.090 & 0.598 & 17.51 & 13.23 & 5.99* & 2.66* & -0.108* \\
 2.095 & 0.650 & 18.18 & 13.53 & 5.68* & 2.42* & -0.122* \\
 2.096 & 0.690 & 18.90 & 13.77 & 5.37* & 2.18* & -0.125* \\
 2.097 & 0.698 & 19.09 & 13.82 & 5.29* & 2.12* & -0.124*
 \\ \hline
 3.995 & 0. & 13.72 & 23.97 & 5.10 & 5.10 & 0 \\
 4.012 & 0.178 & 14.23 & 21.83 & 5.81 & 5.81 & 0.144 \\
 4.029 & 0.280 & 14.69 & 20.78 & 6.24 & 6.24 & 0.240 \\
 4.051 & 0.375 & 15.24 & 19.98 & 6.62 & 6.62 & 0.327 \\
 4.074 & 0.458 & 15.87 & 19.45 & 6.91 & 6.91 & 0.390 \\
 4.098 & 0.528 & 16.53 & 19.14 & 7.11 & 7.11 & 0.426 \\
 4.120 & 0.588 & 17.24 & 18.99 & 7.24 & 7.24 & 0.438 \\
 4.139 & 0.635 & 17.95 & 18.95 & 7.30 & 7.30 & 0.434 \\
 4.160 & 0.682 & 18.93 & 18.98 & 7.34 & 7.34 & 0.417 \\
 4.167 & 0.700 & 19.45 & 19.01 & 7.35 & 7.35 & 0.407
 \\ \hline
 4.321 & 0.478 & 14.90 & 19.77 & 6.89 & 6.89 & 0.496 \\
 4.321 & 0.479 & 15.01 & 19.80 & 6.88 & 6.88 & 0.490 \\
 4.324 & 0.483 & 15.16 & 19.80 & 6.88 & 6.88 & 0.489 \\
 4.325 & 0.489 & 15.29 & 19.79 & 6.89 & 6.89 & 0.490 \\
 4.333 & 0.505 & 15.55 & 19.73 & 6.93 & 6.93 & 0.498 \\
 4.355 & 0.555 & 16.29 & 19.53 & 7.07 & 7.07 & 0.525 \\
 4.377 & 0.602 & 16.99 & 19.40 & 7.18 & 7.18 & 0.541 \\
 4.396 & 0.641 & 17.68 & 19.35 & 7.25 & 7.25 & 0.544 \\
 4.418 & 0.684 & 18.67 & 19.35 & 7.30 & 7.30 & 0.535 \\
 4.420 & 0.686 & 18.74 & 19.36 & 7.30 & 7.30 & 0.534 \\ \hline
\end{tabular}
\end{table}

\begin{table*}
 \centering
 \begin{minipage}{1.0\textwidth}
  \caption{Fitting results for FPS EOS. The relative differences between the fitted moments
  and the actual moments of the space-time used are given as a percentage.}
  \centering
\begin{tabular}{|l|l|r|c|l|r|c|l|r|c|l|}\hline
       &  &\multicolumn{3}{|c|}{$\Omega$: 1-3 kHz} & \multicolumn{3}{|c|}{$\Omega$: 0.3-3 kHz }& \multicolumn{3}{|c|}{$\Omega$: 0.1-3 kHz} \\ \noalign{\smallskip}
 M &  j &  $\frac{\Delta M}{M}$ & $\frac{\Delta J}{J}$ & $\frac{\Delta M_2}{M_2}$
 &
 $\frac{\Delta M}{M}$ & $\frac{\Delta J}{J}$ & $\frac{\Delta M_2}{M_2}$
 &
 $\frac{\Delta M}{M}$ & $\frac{\Delta J}{J}$ & $\frac{\Delta M_2}{M_2}$ \\
  (km)&   & (per cent) & (per cent) & (per cent) & (per cent) & (per cent) & (per cent) & (per cent) & (per cent) & (per cent) \\ \hline
 2.067 & 0. & 0.777 & - & - & 0.211 & - & - & 0.1025 & - &  - \\
 2.071 & 0.201 & 0.431 & 14.63 & 154.3 & 0.103 & 4.228 & 54.34 & 0.0564 & 2.562 & 36.38 \\
 2.077 & 0.301 & 0.285 & 6.650 & 47.46 & 0.074 & 2.058 & 17.50 & 0.0390 & 1.191 & 11.20 \\
 2.083 & 0.385 & 0.269 & 4.914 & 26.95 & 0.067 & 1.434 & 9.310 & 0.0339 & 0.804 & 5.750 \\
 2.087 & 0.452 & 0.253 & 3.923 & 18.01 & 0.059 & 1.070 & 5.753 & 0.0289 & 0.576 & 3.388 \\
 2.093 & 0.507 & 0.236 & 3.278 & 13.26 & 0.051 & 0.830 & 3.872 & 0.0242 & 0.424 & 2.141 \\
 2.098 & 0.557 & 0.219 & 2.777 & 10.10 & 0.044 & 0.641 & 2.625 & 0.0196 & 0.305 & 1.317 \\
 2.102 & 0.603 & 0.260 & 3.020 & 9.940 & 0.053 & 0.707 & 2.590 & 0.0241 & 0.340 & 1.300 \\
 2.106 & 0.636 & 0.301 & 3.275 & 9.995 & 0.047 & 0.585 & 1.957 & 0.0201 & 0.262 & 0.876 \\
 2.109 & 0.666 & 0.285 & 2.968 & 8.561 & 0.041 & 0.479 & 1.452 & 0.0163 & 0.194 & 0.538
 \\\hline
 2.658 & 0. & 2.236 & - & - & 0.587 & - & - & 0.3116 & - & -  \\
 2.664 & 0.163 & 1.277 & 48.71 & 1314. & 0.306 & 14.07 & 463.3 & 0.1537 & 7.982 & 295.4 \\
 2.674 & 0.260 & 0.723 & 17.67 & 285.5 & 0.185 & 5.359 & 104.2 & 0.0867 & 2.853 & 62.76 \\
 2.686 & 0.349 & 0.479 & 8.802 & 99.28 & 0.133 & 2.838 & 37.66 & 0.0585 & 1.415 & 21.26 \\
 2.701 & 0.436 & 0.358 & 5.231 & 43.28 & 0.069 & 1.192 & 11.70 & 0.0339 & 0.639 & 6.825 \\
 2.714 & 0.500 & 0.329 & 4.124 & 27.50 & 0.055 & 0.801 & 6.068 & 0.0248 & 0.384 & 3.029 \\
 2.727 & 0.562 & 0.237 & 2.643 & 14.56 & 0.039 & 0.477 & 2.593 & 0.0144 & 0.171 & 0.701 \\
 2.736 & 0.602 & 0.214 & 2.185 & 10.47 & 0.028 & 0.285 & 0.977 & 0.0069 & 0.044 & 0.373 \\
 2.744 & 0.633 & 0.194 & 1.859 & 7.961 & 0.019 & 0.148 & 0.005 & 0.0008 & 0.047 & 1.023 \\
 2.750 & 0.654 & 0.179 & 1.641 & 6.454 & 0.012 & 0.056 & 0.590 & 0.0035 & 0.108 & 1.409
 \\\hline
 2.823 & 0.427 & 0.351 & 5.143 & 56.62 & 0.066 & 1.146 & 14.91 & 0.0322 & 0.606 & 8.542 \\
 2.823 & 0.428 & 0.439 & 6.348 & 67.69 & 0.089 & 1.528 & 19.39 & 0.0455 & 0.849 & 11.73 \\
 2.825 & 0.432 & 0.350 & 5.064 & 52.94 & 0.066 & 1.119 & 13.80 & 0.0318 & 0.589 & 7.859 \\
 2.826 & 0.439 & 0.348 & 4.942 & 49.69 & 0.064 & 1.077 & 12.73 & 0.0309 & 0.561 & 7.159 \\
 2.829 & 0.450 & 0.343 & 4.754 & 45.45 & 0.062 & 1.010 & 11.28 & 0.0295 & 0.518 & 6.204 \\
 2.840 & 0.492 & 0.322 & 4.027 & 31.97 & 0.052 & 0.749 & 6.659 & 0.0228 & 0.346 & 3.137 \\
 2.856 & 0.552 & 0.231 & 2.551 & 16.16 & 0.036 & 0.419 & 2.439 & 0.0119 & 0.127 & 0.338 \\
 2.871 & 0.609 & 0.195 & 1.895 & 9.463 & 0.018 & 0.139 & 0.140 & 0.0002 & 0.059 & 1.356 \\
 2.882 & 0.647 & 0.116 & 0.996 & 3.692 & 0.007 & 0.174 & 2.234 & 0.0141 & 0.247 & 2.650 \\
 2.884 & 0.658 & 0.108 & 0.893 & 2.977 & 0.010 & 0.218 & 2.478 & 0.0164 & 0.276 & 2.799 \\ \hline
\end{tabular}
\end{minipage}
\end{table*}

\subsection{Neutron star models}

In order to construct the analytic space-time exterior to a
compact object, one has to choose the appropriate multipole
moments. For neutron stars, these moments can be read from
numerical models constructed by assuming realistic equations of
state. There are several schemes developed for numerically
integrating stellar models (see \cite{Sterg}, and for an extended
list of numerical schemes see \cite{Lrr}). We have used
Stergioulas's RNS code for the construction of the models.

The parameters used to construct the analytic space-time models,
i.e., $M,\;a,\;k$ and $b$,  are evaluated from the first four
multipole moments ($M,\;J,\;M_2,\;S_3$) as given by equations
(\ref{moments}). In order to cover more space on the
``neutron-star parameter space", these moments are obtained from
numerically calculated neutron stars with EOSs of varying
stiffness. For that purpose we have chosen AU as a typical soft
EOS, FPS as a representative moderate stiff EOS and L to describe
stiff EOS. After the numerical models are computed, the multipole
moments are evaluated according to the algorithm described in
\cite{numericalmoments} since there is a systematic discrepancy in
the methods used up to now in the literature.

For the specifics of the various models chosen, we have followed
\cite{BertiSterg}. Thus, we have constructed the same constant
rest-mass sequences as the ones presented in \cite{BertiSterg} for
the corresponding equations of state. That is, every sequence is
comprised of models with the same baryon number but different
rotation. Specifically for every equation of state, 3 sequences of
10 models with varying rotations are constructed, which correspond
to:

\begin{enumerate}
\item a sequence terminating to a neutron star of $1.4
M_{\odot}$ in the non-rotating limit,
\item a sequence terminating at the maximum-mass neutron star in
the non-rotating limit,
\item a supermassive sequence that does not terminate at a
non-rotating model.
\end{enumerate}

All the sequences end at the mass-shedding limit on the side of
fast rotation, i.e., at the limit where the angular velocity of a
particle at the equator is equal to the Keplerian velocity at that
radius. These sequences are the so called {\it evolutionary
sequences} in the sense that an isolated neutron star (i.e. with
constant number of baryons) that is losing angular momentum will
evolve along such a sequence. In the case that there is accretion
onto the neutron star, the baryon number is not constant and the
exact evolution of the neutron star depends on several parameters.
For a LMXB though, where the mass of the companion is
$<1M_{\odot}$, the overall mass transfer is probably a small
fraction of a solar mass and thus a neutron star (with mass
$\geqslant 1.4M_{\odot}$) would evolve following approximately
such a sequence. The choice of these particular sequences was made
so that the parameter space of the equilibrium neutron star models
would be fully sampled (for details on neutron star models and the
evolutionary sequences, see \cite{CST}).

Once all the necessary parameters are at hand the various
frequencies can be easily calculated. As an example, we present in
Table 1 the radii of the ISCO, the orbital frequency $\Omega$ and
the two precession frequencies $\Omega_{\rho}$ and $\Omega_z$ at
that radius, for the 3 sequences constructed with L EOS. We also
present the masses, the spin parameters ($j=J/M^2$) and the
surface radii of the corresponding models. In some models the
surface of the star overcomes the radius of the marginally stable
orbit and in these cases the respective frequencies are computed
on the radius of the surface. One notices that the periastron
precession frequency $\Omega_{\rho}$ coincides with the orbital
frequency $\Omega$ at the ISCO and that for the non-rotating
models, the nodal precession frequency $\Omega_z$ is zero, as
expected. The ISCO radii are calculated from the analytic metric,
i.e., the two-soliton metric, and thus deviate from the ones given
by Stergioulas's code by at most 3 per cent
(\cite{Pappas2,Pappas3}).

\section{Fitting the frequencies using the\\ * asymptotic expressions}

In several occasions, there have been observed twin kHz QPOs with
the upper QPO being in a relatively low frequency range (see eg.
\cite{boutloukos}) and could thus be considered to come from a
region of the accretion disc further than the ISCO (depending on
the mass of the central object). In these cases one could directly
apply the asymptotic expression presented by Ryan and try to fit
for the multipole moments.

In order to explore whether this method could provide useful
measurements of the first moments, we will use simulated data
[pairs of $(\Omega,\,\Omega_{\rho})$] for the ``observed"
frequencies of QPOs constructed by assuming the two-soliton
analytic solution. For every numerical model of those discussed in
the previous section with multipole moments $M,\;J,\;M_2,\;S_3$, a
corresponding two-soliton space-time is constructed and the
orbital and precession frequencies for various radii are
calculated. From these frequencies, we select the
$(\Omega,\,\Omega_{\rho})$ pairs and plot the ratio
$\Omega_{\rho}/\Omega$ as a function of $\Omega$. We then fit
these data with the expression (\ref{ryan1}) and calculate the
coefficients. From these coefficients, the fitted moments are
calculated and compared to the ones used initially to construct
the two-soliton space-time.

For the fits we will use data points which correspond to orbital
frequencies lower than 3 kHz, which depending on the specifics of
the model (mass, rotation and higher moments) correspond to
distances greater than 1.4-2.5~$R_{isco}$. Specifically we will
use three frequency ranges for the fits, i.e., from 3 kHz down to
1 kHz (which is between 1.4-5~$R_{isco}$ depending on the model),
3-0.3 kHz (1.4-10~$R_{isco}$) and 3-0.1 kHz (1.4-20~$R_{isco}$).
As it will be discussed in the following, the fit in the moments
improves as we increase the frequency range, so these ranges have
been chosen to demonstrate that effect.

QPOs are observed in the spectrum of X-ray photons in the range of
2-60 keV (see e.g. \cite{derKlis,lamb}), i.e., within a couple of
orders of magnitude in energy. Generally in an accretion disc, the
temperature of the disc decreases with the distance as $\propto
(r/R_{isco})^{-3/4}$ (see e.g. \cite{krolik}) which means that in
the range between $\sim(1-20)\,R_{isco}$ there would be a decrease
in emitted photons energy of an order of magnitude. Thus, although
the $20\,R_{isco}$ is not a limiting distance in any sense for the
disc or the QPOs, it is a reasonable choice for a final fitting
range for our purpose.

The results of the fits for the models that are based on the FPS
EOS are presented in Table 2. These results are typical for all
three EOSs. One sees from Table 2 that in the range of frequencies
between 1 and 3 kHz (which correspond to distances up to
$5R_{isco}$), the mass can be very accurately constrained. The
same applies for the angular momentum for most of the models and
especially for those with faster rotation. The quadrupole is
generally not very well constrained but for the cases of fast
rotation, one could have accuracies better than 30 per cent. As
the fitting range increases, the accuracy of the fits improves and
becomes quite good for the quadrupole as well. One should also
note that the accuracy of a fit deteriorates with increasing mass.
That is because the higher the mass of the model, the closer we
get to the ISCO for the selected frequency ranges.

\section{Fitting the frequencies by using\\ * templates}
\begin{figure}

  \includegraphics[width=0.5\textwidth]{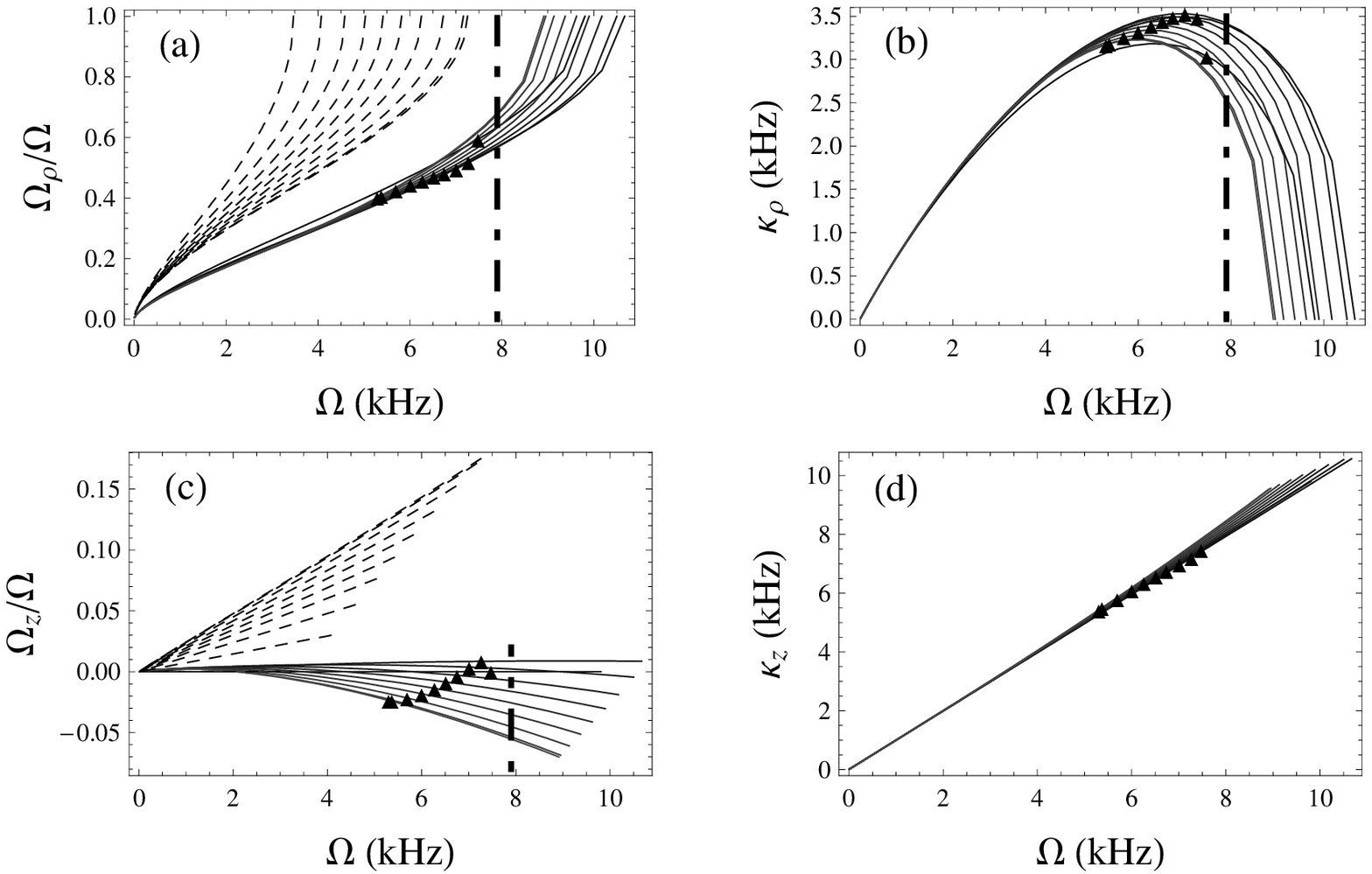}
  \includegraphics[width=0.5\textwidth]{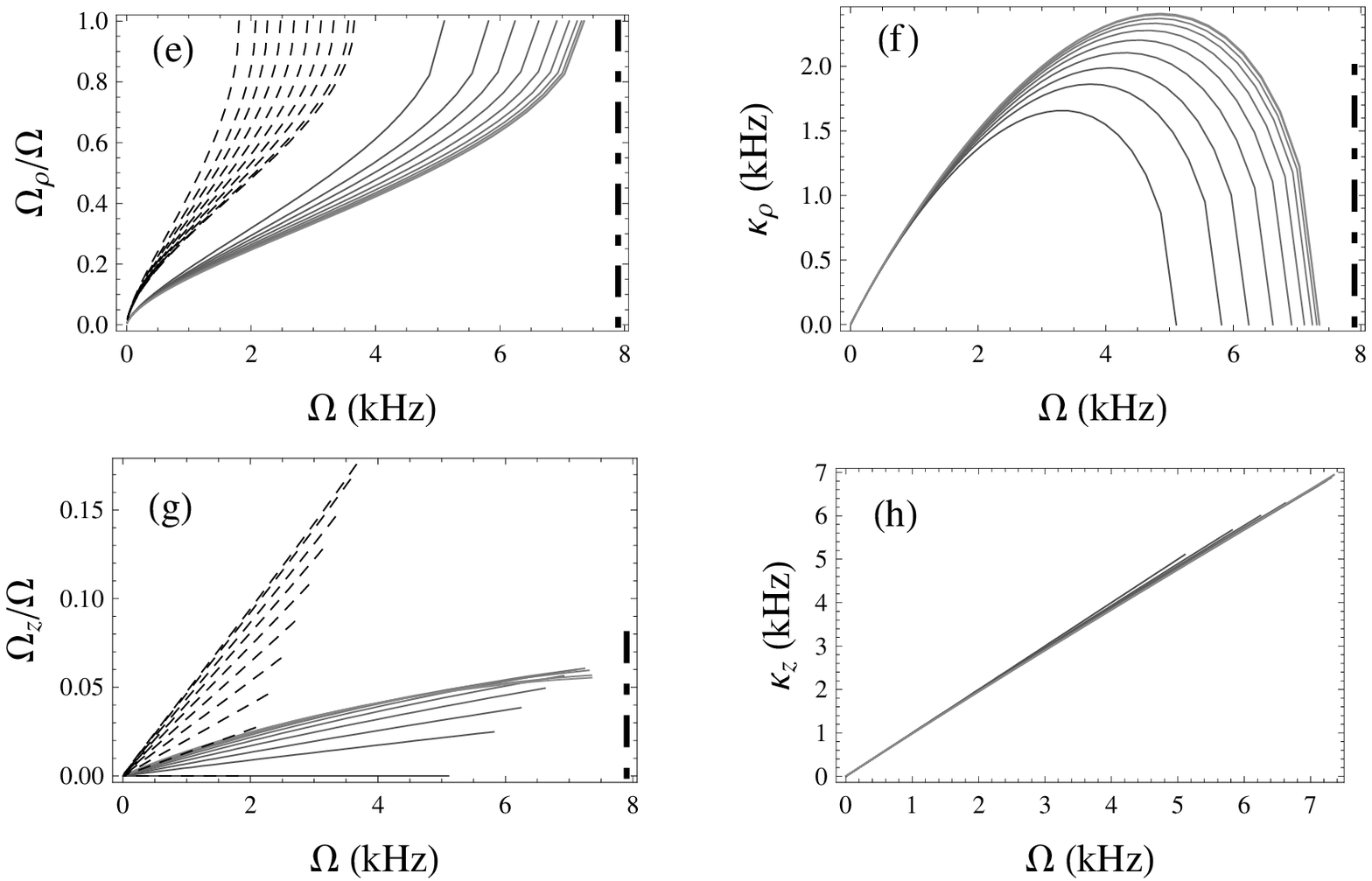}
  \includegraphics[width=0.5\textwidth]{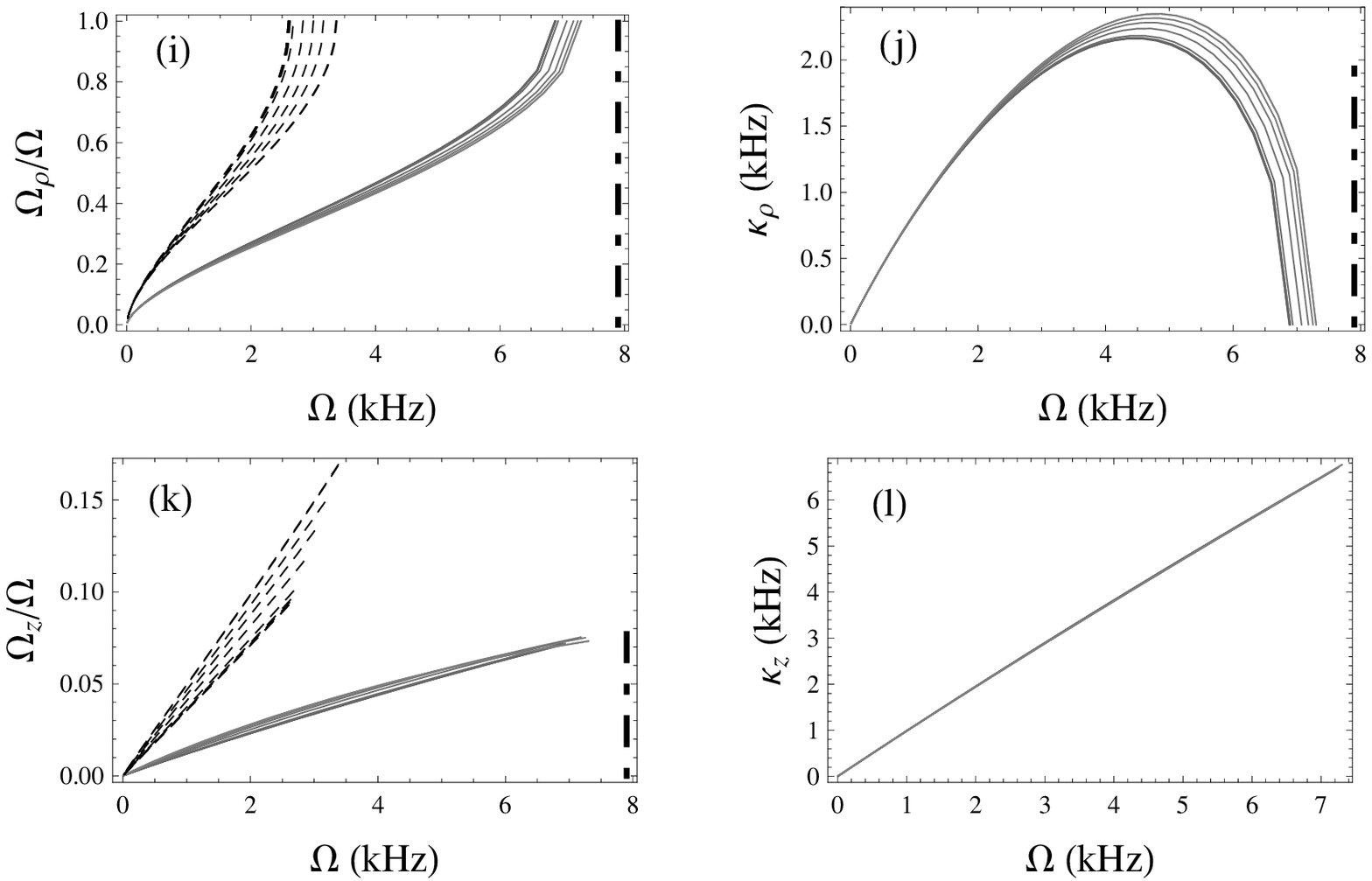}
  \caption{Plot of frequencies for the three sequences of models for the L EOS.
  Plots (a-d) are for sequence (i), (e-h) are for sequence (ii) and (i-l) are for
  sequence (iii). All curves are terminated at the ISCO. The black triangles represent
  the surface of the neutron star for the particular model, if the ISCO is under the
  surface. The dashed lines correspond to the frequencies for Kerr space-times of the same mass and
  rotation parameter as the corresponding neutron star models. The dashed-dotted line denotes the maximum
  QPO frequency measured up to now.}

\end{figure}

Since Ryan's expressions are asymptotic expansions in
$\upsilon=(M\Omega)^{1/3}$, one should not expect them to be very
precise in giving the relationship between the frequencies in the
region of the space-time near the marginally stable orbit, which
corresponds to $\Omega$ frequencies, greater than about 3-4 kHz.
For this reason, it would be useful to have templates of
frequencies, for different models constructed with different EOSs,
on which to project the observations. In this section we will
present such templates using the three EOSs that we have already
discussed.

For every EOS, we produce by means of the two-soliton metric, the
three frequencies $\Omega,\,\Omega_{\rho}$ and $\Omega_z$ and the
two oscillation frequencies, $\kappa_{\rho}=\Omega-\Omega_{\rho}$
for the radial perturbation and $\kappa_z=\Omega-\Omega_z$ for the
vertical perturbation. The precession frequencies, actually the
ratios $\Omega_{\rho}/\Omega,\,\Omega_z/\Omega$, and oscillation
frequencies are plotted against the orbital frequency $\Omega$ for
the three sequences of neutron star models. In Fig. 1 we present
the plots for the L EOS. The behavior of the frequencies is
similar for all the equations of state.

As it can be seen from the templates, in the case of the orbital
frequency and the periastron precession frequency, as the orbit
approaches the marginally stable orbit the two frequencies become
equal. That is to be expected, since the precession frequency is
the difference between the orbital frequency and the frequency of
the radial perturbation, which goes to zero at the ISCO (the
change of sign from positive to negative of the square of the
frequency of the radial perturbation marks the onset of the radial
instability). That could be used as a criterion of whether a pair
of observed kHz QPOs are the orbital and the precession
frequencies or not. If they are, then one would expect that the
evolution of the two frequencies should bring them together on the
ISCO before they disappear.

Another important feature that arises is that the precession
frequencies behave differently than the corresponding Kerr
frequencies. More specifically, in some cases the nodal precession
at high orbital frequencies (which correspond to the inner part of
the disc near the marginally stable orbit) becomes zero. That is
because the vertical perturbation of the orbit has a frequency
equal to the orbital one. That never happens in the Kerr geometry
(see Fig. 1) and the effect is due to the difference between the
multipole moments of a realistic neutron star and those of the
Kerr space-time from the quadrupole and higher. It has been shown
by \cite{poisson} that while the Kerr quadrupole is equal to minus
the square of the spin parameter, $q=-j^2$, where $q=M_2/M^3$, for
realistic neutron stars the quadrupole becomes proportional to
that of Kerr's with a proportionality constant greater than 1 and
increasing with increasing stiffness of the EOS (it also decreases
at higher masses). Similar behavior is observed for the spin
octupole as well (discussed in \cite{numericalmoments}). From that
perspective a Kerr black hole behaves as an object with a
super-soft EOS.

Going further towards the inner regions of the disc (closer to the
ISCO), the frequency of the vertical perturbation becomes greater
than the orbital frequency up until the ISCO. This is evident in
the plot of $\Omega_z/\Omega$ versus $\Omega$ in the case of the L
EOS [for the models of the sequence (i) of the corresponding
neutron stars]. That effect would produce a distinctive behavior
in the observed QPOs that could be used to select an equation of
state, since whether or not such an effect is present or
specifically the frequency range where one would expect to see it,
depends on the stiffness of the EOS of the neutron star.

We can see similar behavior in the corresponding plots for the FPS
and AU EOSs, with the difference that the vertical oscillation
frequency does not always become greater than the orbital
frequency before the ISCO is reached, as can be seen in Figs 2 and
3. The distinctive drop of the ratio $\Omega_z/\Omega$ though
remains.

In the plots we also indicate by a small triangle the surface of
the star whenever it is at a radius that is further than the ISCO.
Generally the ISCO is swallowed by the star for models of the
first sequence for every EOS. For these models, the decrease of
$\Omega_z/\Omega$ is also more prominent, thus it becomes apparent
even if the ISCO is swallowed by the star. As it can be seen for
the case of L EOS, it might become negative even before reaching
the star's surface.

Finally, a dashed-dotted line marks the position of the maximum
measured QPO frequency ($\nu=1258\, \mathrm{Hz}$) which
corresponds to $\Omega\sim 7.9\,\mathrm{kHz}$. The plots show
something that has been discussed in the literature, namely the
fact that this particular frequency is incompatible with the L
equation of state, since there is no model that could produce such
a high frequency either because of the position of the ISCO or
because of the position of the surface.

\begin{figure}

  \includegraphics[width=0.5\textwidth]{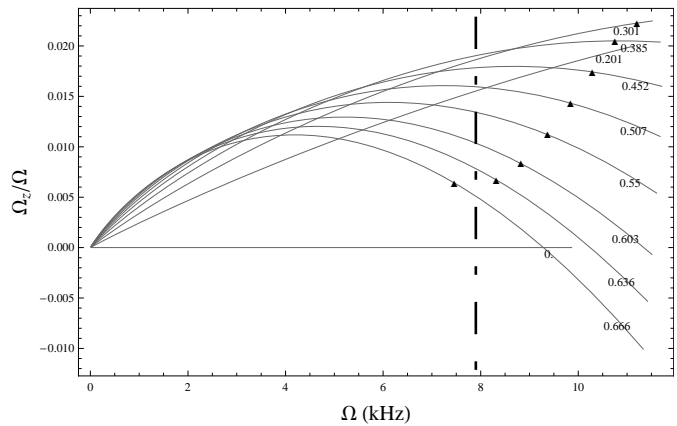}
  \caption{$\Omega_z/\Omega$ as a function of $\Omega$ for the FPS EOS. The
  spin parameters $j=J/M^2$ are also indicated for every model.}

\end{figure}

\begin{figure}

  \includegraphics[width=0.5\textwidth]{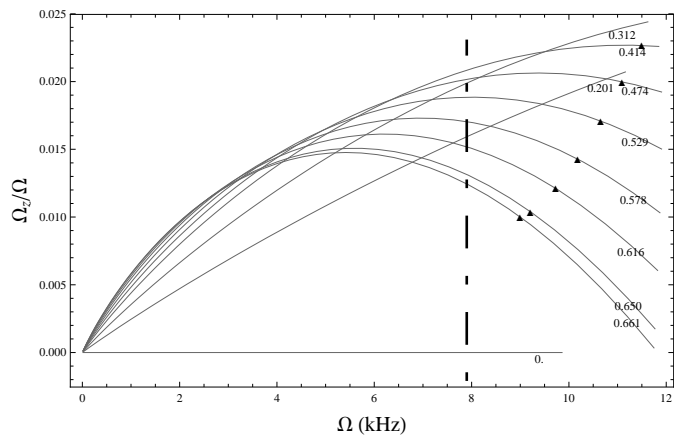}
  \caption{Same as Fig. 2 for the AU EOS.}

\end{figure}

\section{Distinguishing different\\ * equations of state of Neutron Stars}

In this section, we will use the templates presented in the
previous section and we will attempt to draw some conclusions
about the EOS from observations. For illustrative purposes we will
use observations from two sources as well, i.e., Scorpius X-1 (Sco
X-1) represented by the filled squares and Circinus X-1 (Circ X-1)
represented by the empty squares, as obtained from
\cite{derKlisSco} and \cite{boutloukos}.

\begin{figure}

  \includegraphics[width=0.5\textwidth]{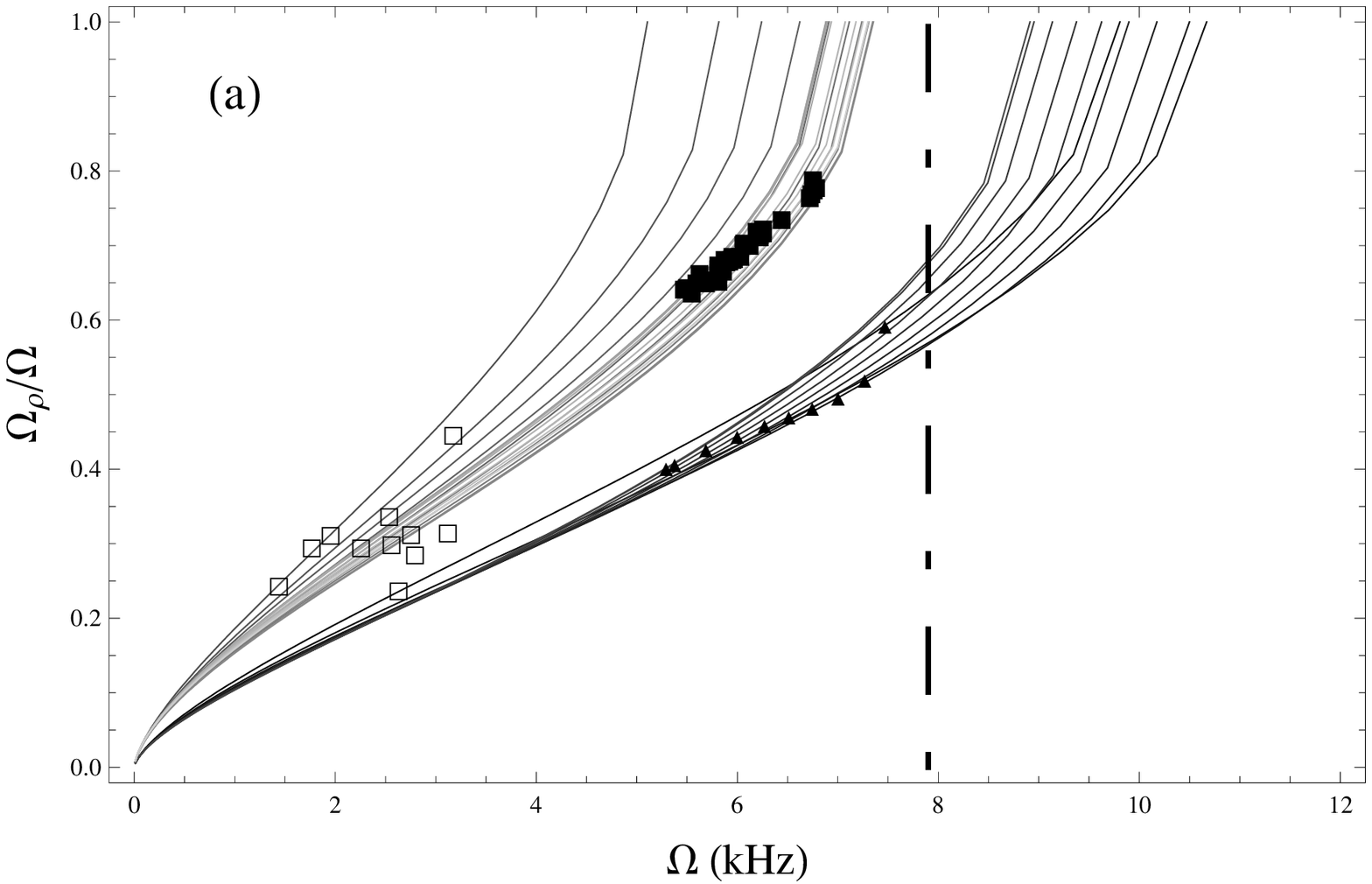}
  \includegraphics[width=0.5\textwidth]{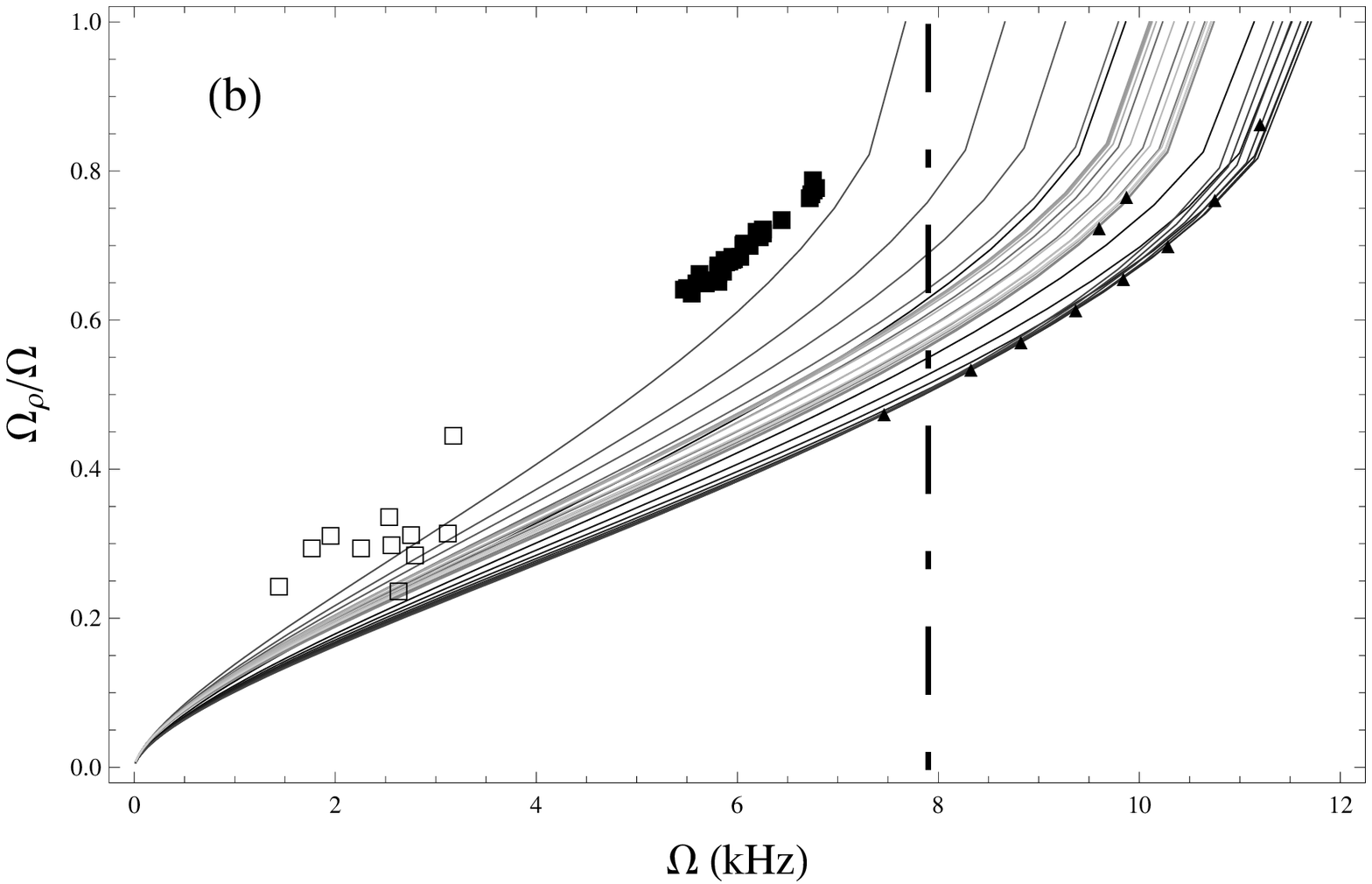}
  \includegraphics[width=0.5\textwidth]{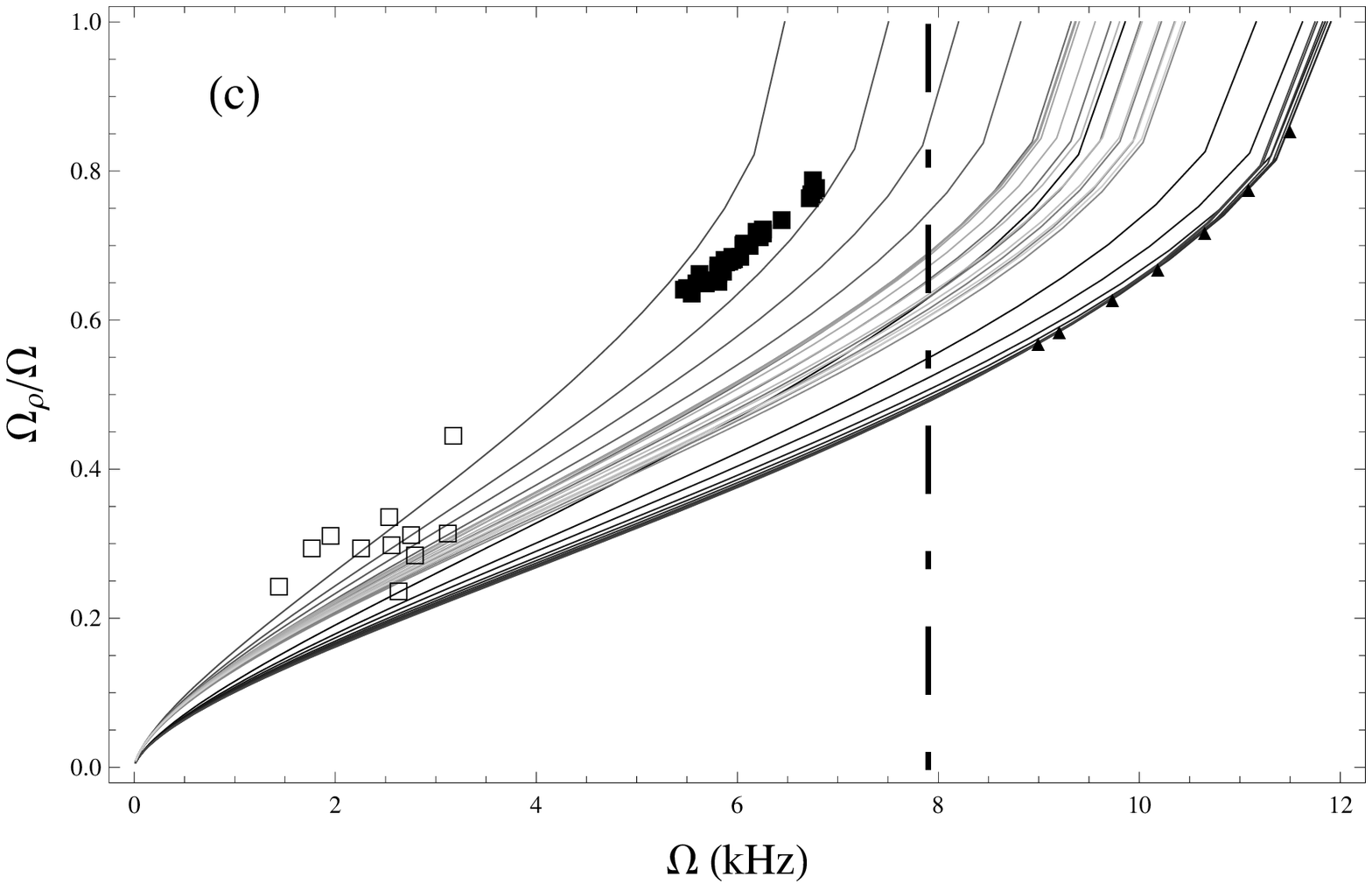}
  \caption{QPOs observations compared with the templates for the equations
  of state (a) L, (b) FPS, and (c) AU. The black triangles
  indicate the surface whenever it is present. The filled squares are observations
  related to the source Sco X-1 and the empty squares are observations of
  Circ X-1.}

\end{figure}

\begin{figure}

  \includegraphics[width=0.5\textwidth]{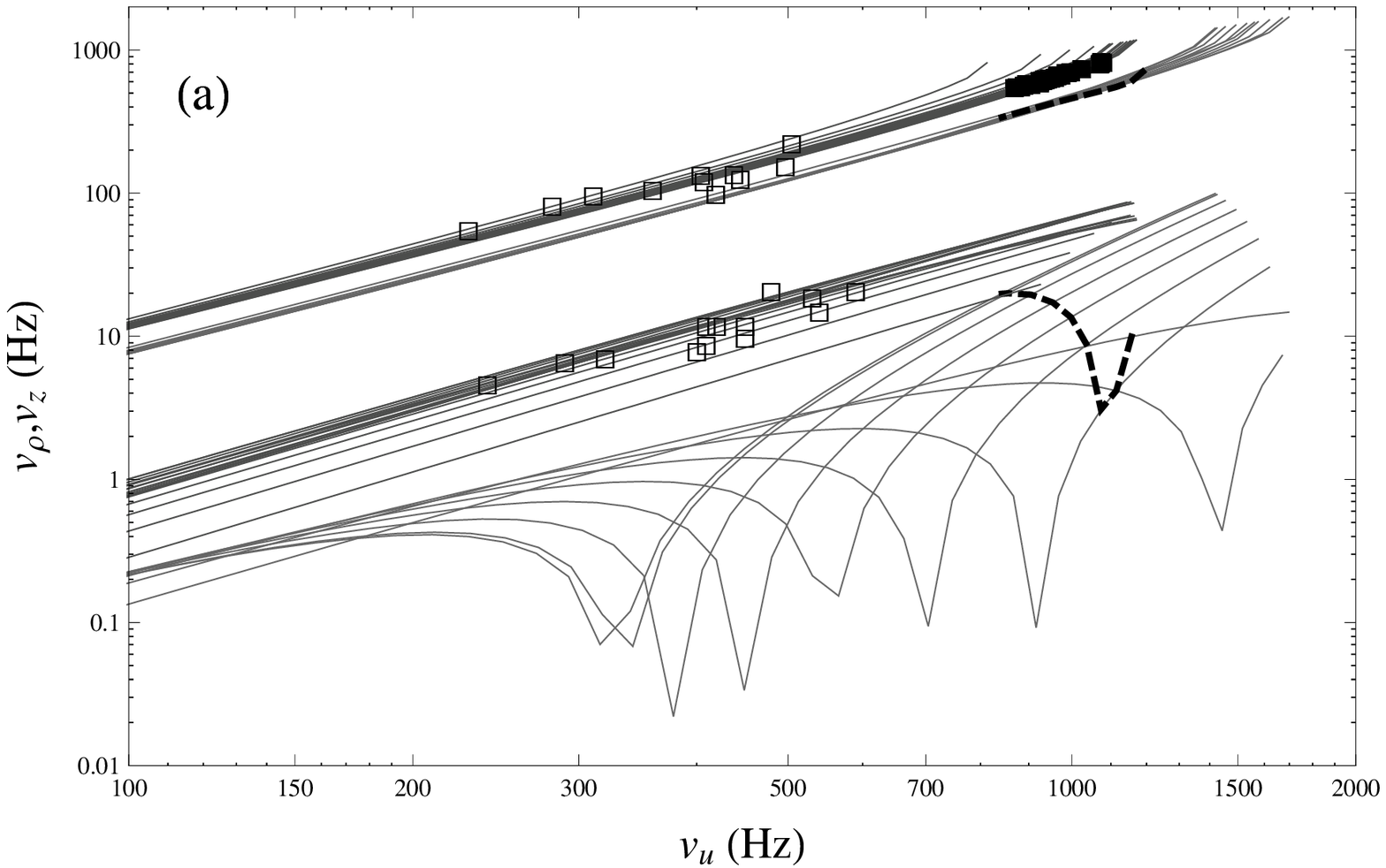}
  \includegraphics[width=0.5\textwidth]{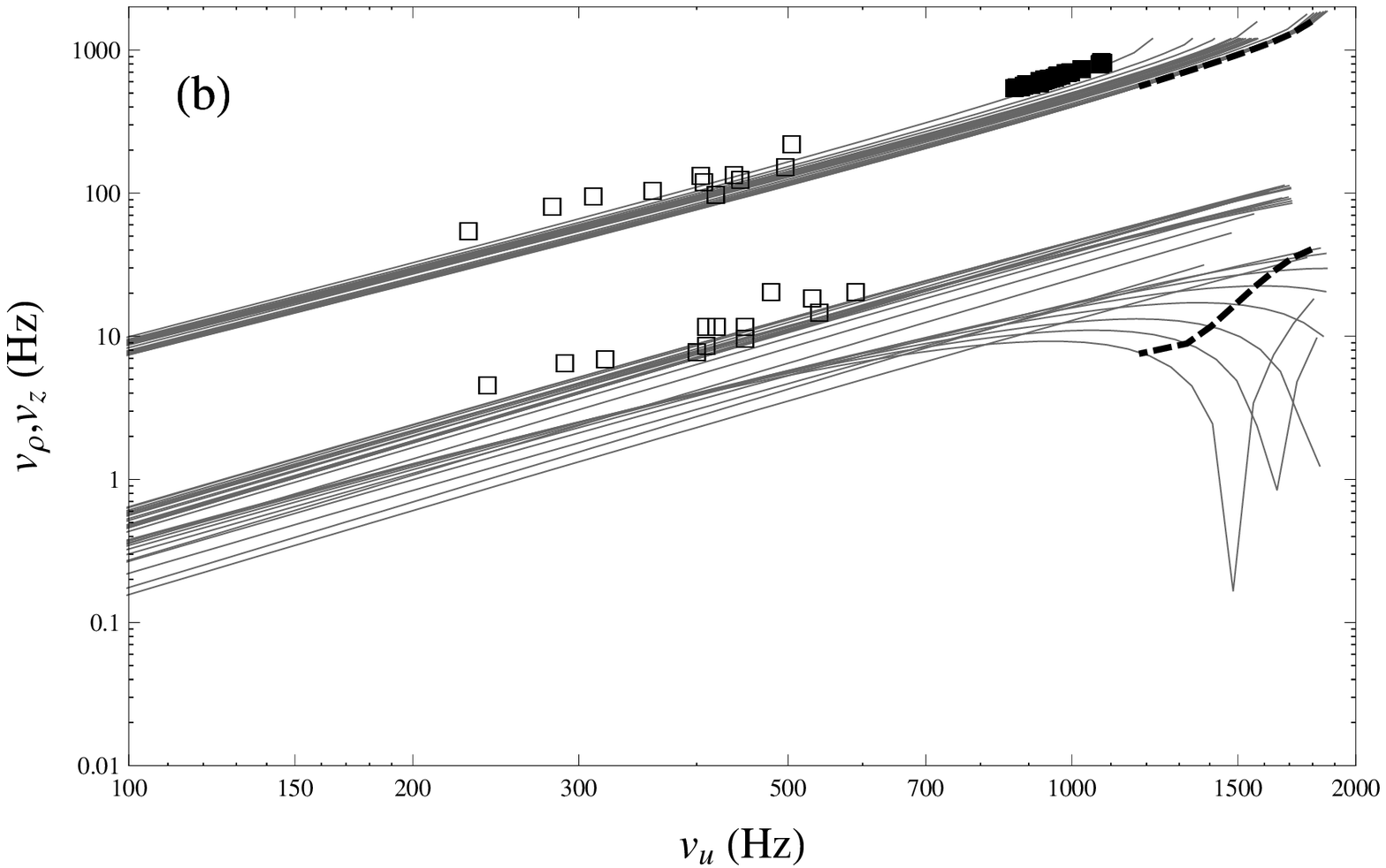}
  \includegraphics[width=0.5\textwidth]{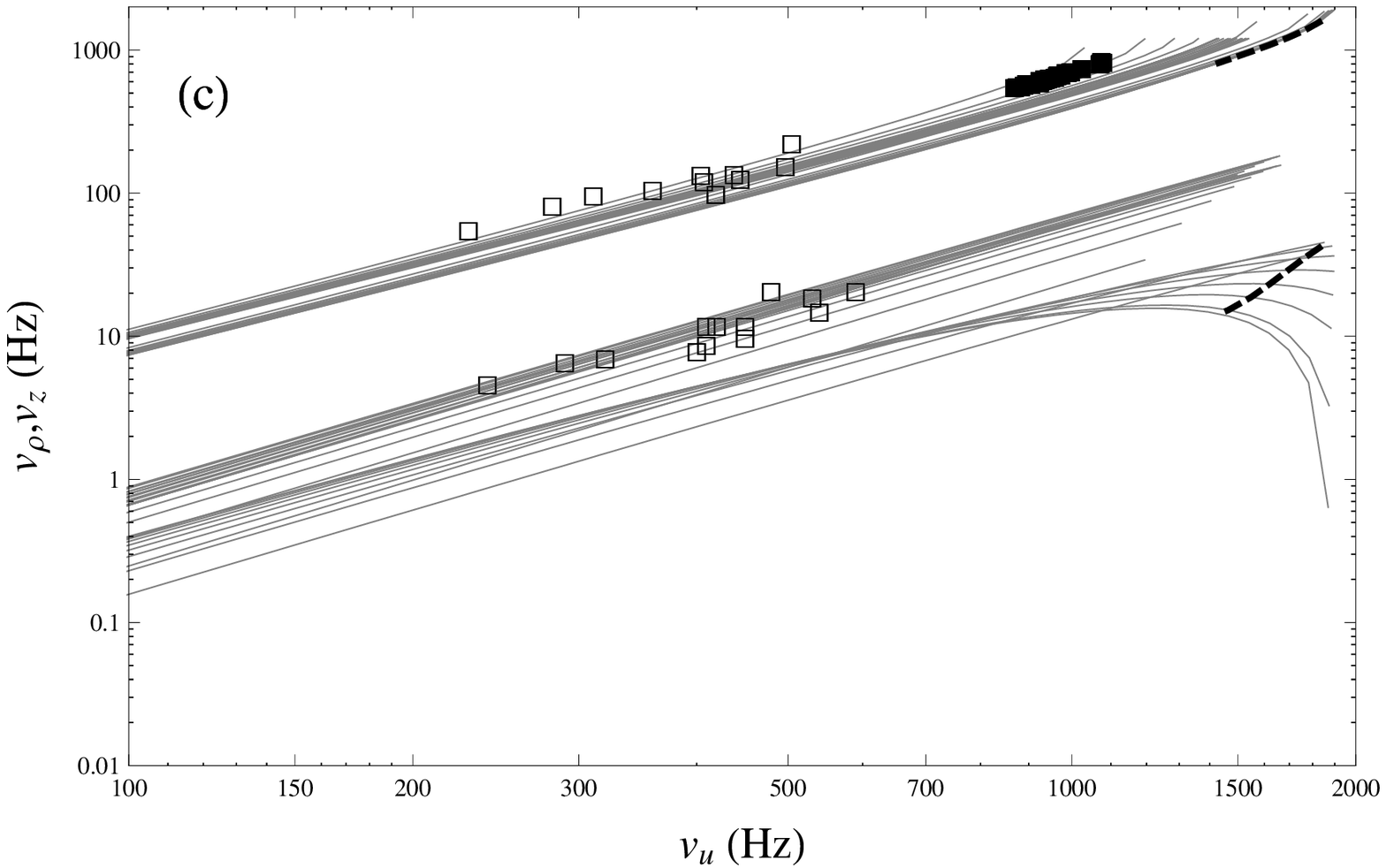}
  \caption{Simulated QPO frequencies related to the periastron and nodal precession
  ($\nu_{\rho},\;\nu_{z}$) as functions of the orbital frequency ($\nu_{u}$)
  for the models with equations of state (a) L, (b) FPS, and (c) AU. Since $\Omega_z$ could be
  either positive or negative, the $\nu_{z}$ is simply $|\Omega_z/2\pi|$. The characteristic
  features that appear on the nodal precession frequencies of the neutron stars do not appear in the
  corresponding frequencies for a Kerr background.}

\end{figure}

Once we have the frequency templates for the various models that
correspond to different EOSs, we can compare them to observations
of the orbital frequency and of the periastron precession
frequency, i.e., the twin kHz QPOs. Such a comparison is shown in
Fig. 4, where we have plotted the curves of $\Omega_{\rho}/\Omega$
as a function of $\Omega$ for all the models of the three EOSs.
The plots also show the observations of Sco X-1 and Circ X-1. An
appropriate selection of observations of several sources compared
with the frequency templates could select or exclude EOSs. For
example the two sources shown here, appear not to fit very well
with the FPS EOS. Furthermore, if one had good observations on one
source, then one could select a specific model and thus estimate
the mass and the spin of the neutron star, as well as its higher
moments. That would offer the opportunity for investigating the
interior structure of a neutron star.

In Fig. 5 we have plotted for the different EOSs what would be the
observed frequencies $\nu$ of the various models, assuming that
the upper kHz QPOs correspond to the orbital frequency, the lower
kHz QPOs correspond to the periastron precession and the
low-frequency QPOs correspond to the nodal precession. The plots
also indicate with a dashed curve the surface for the models that
have their ISCO inside the star. This type of plots are often used
to present data from various sources. For illustrative purposes we
also show with black squares the kHz QPOs of Sco X-1 and with
empty squares the kHz (upper grouping) and low-frequency (lower
grouping) QPOs of Circ X-1. Although such a plot ``masks" the
differences of each source when we are considering kHz QPOs, it
makes evident the behavior of the frequency of the nodal
precession (low-frequency QPOs), that we discussed in the previous
section. The turning of the low-frequency QPO to a horizontal path
and the subsequent drop of the curve as the upper kHz QPO
increases, that is evident in the L EOS plots, would be a clear
signature that the observed frequencies are indeed related to
orbital motions. The range of frequencies that such a behavior
would be observed would also distinguish between different EOSs,
as the three plots show, while with sufficient resolution one
could identify a specific model from the frequency of that
``drop".

It should also be pointed out that the three different sequences
of neutron star models that we have used group together in a
similar fashion. The first sequence that corresponds to neutron
stars that are not near the maximum stable mass at the non
rotating limit ($M=1.4M_{\odot}$) exhibits lower low-frequency
QPOs that present the characteristic ``drop" discussed previously,
while the other two sequences (the one at the maximum stable mass
in the non-rotating limit and the one with no stable non-rotating
limit) group together to higher low-frequencies and do not show
the aforementioned ``drop". Differences between the three
sequences for the kHz QPOs are not that prominent, but depending
on the EOS one could detect a small gap between the first sequence
and the other two.

\section{Conclusions}

We have investigated the merits of applying some new tools in the
study of compact objects from the observation of QPOs in the X-ray
flux of accreting systems.

We have shown that Ryan's asymptotic expressions for the
frequencies, combined with the appropriate observations, can be
useful in constraining the first three multipole moments of the
compact object. That would offer us the opportunity to probe the
structure of neutron stars, as well as the possibility of testing
the validity of the no-hair theorem for black holes.

We have also demonstrated the potential benefits of using analytic
space-times that are constructed so as to be faithful to the
geometry around realistic neutron star models. The comparison of
the observed QPOs to the ones calculated with such space-times,
could potentially select or exclude EOSs for neutron stars or
could even select specific neutron star models for specific
sources. We have also shown that there are characteristic
properties of the frequencies of the orbits in realistic neutron
star space-times that the Kerr and Schwarzschild geometries cannot
capture. Specifically the way that the nodal precession varies as
one goes from lower to higher orbital frequencies would be a
signature of the orbital nature for any observed QPOs that exhibit
that behavior and could be used to distinguish different EOSs and
probably even particular models of neutron stars.

Following this line of research it would be worth pursuing
further, with the help of analytic space-times, the investigation
of orbits that deviate significantly from the equatorial plane as
well as the investigation of possible orbital resonances that can
be observed for these space-times. The use of analytic instead of
numerical geometries would greatly simplify such an exploration.

\section*{Acknowledgments}

I would like to thank Haris Apostolatos, Nektarios Vlahakis,
Apostolos Mastichiadis, Kostas Glampedakis and Stratos Boutloukos
for useful discussions. I would also like to thank Kostas Kokkotas
for his hospitality at the Theoretical Astrophysics Section in
T$\ddot{u}$bingen (TAT) and I thank Nikos Stergioulas for
providing me access to his RNS numerical code. Finally, I would
like to thank the anonymous referee for his/her comments and
suggestions that helped improve on the clarity in the presentation
of this work. This work has been supported by the IKY-DAAD
programme (IKYDA 2010).

%
%
%
%

\label{lastpage}

\end{document}